\documentstyle[epsfig,preprint,aps]{revtex}
\begin{document}
%\tighten
%\widetext
\draft
\def\bra#1{{\langle #1{\left| \right.}}}
\def\ket#1{{{\left.\right|} #1\rangle}}
\def\bfgreek#1{ \mbox{\boldmath$#1$}}
\title{Medium modification of the nucleon axial form factor}
%\footnotemark{}
%\footnotetext{ADP-97-XX/TYYY}
\author{D. H. Lu$^1$\footnote{dhlu@zimp.zju.edu.cn}, 
A. W. Thomas$^2$\footnote{athomas@physics.adelaide.edu.au},
 and K. Tsushima$^{2,3}$\footnote{tsushima@kn3.physast.uga.edu}}
\address{$^1$Department of Physics and 
         Zhejiang Institute of Modern Physics,\break
         Zhejiang University,
         Hangzhou 310027, China}
\address{$^2$Department of Physics and Mathematical Physics\break
 and
        Special Research Centre for the Subatomic Structure of Matter,\break
        University of Adelaide, Australia 5005}
\address{$^3$Department of Physics and Astronomy,\break
         University of Georgia, Athens, GA 30602, USA}
\maketitle
\vspace{-9.3cm}
\hfill ADP-01/52-T484
\vspace{9.3cm}
\begin{abstract}
We study the modification of the nucleon axial form factor in nuclear matter. 
The internal quark substructure of the nucleon is self-consistently 
described by the quark meson coupling model. 
We find that the axial form factor of the bound nucleon
is quenched considerably from that of the free nucleon.
The axial vector coupling constant, $g_A$, is reduced by roughly 10\% 
at normal nuclear matter density and the axial form factor varies within 8\%  
for moderate momentum transfer.
\end{abstract}
\vspace{0.5cm}
\pacs{PACS numbers: 12.39.Ba, 21.65.+f, 13.40.Gp, 25.70Bc}
%\keywords{quark-meson coupling model, in-medium nucleon, 
%electromagnetic form factors, y-scaling} 
%

%\section{Introduction}

There is strong evidence that hadron properties must
undergo substantial modifications in nuclear medium~\cite{QHD,DBHF,Tony94}.
A number of experiments, such as the variation of nucleon  structure 
functions in lepton deep-inelastic scattering off nuclei 
(the nuclear EMC effect)~\cite{EMC}, 
the quenching of the axial vector coupling constant, $g_A$, 
in nuclear $\beta$-decay~\cite{gA}, and the missing strength 
of the response functions in nuclear quasielastic  electron 
scattering~\cite{quasi}, have stimulated investigations of 
whether or not quark degrees of freedom play any significant role. 
Though the conventional interpretation arising through polarization effects
and other hadronic degrees of freedom ($\Delta$-excitations, meson exchange 
currents, etc.) undoubtedly play a role~\cite{Mulders90,MEC},
it is rather interesting to explore, as well, the possible effects of a
change in the internal structure of the bound nucleon.

The successes of quantum hadrodynamics (QHD) leave little doubt 
that relativistic nuclear 
phenomenology is essential in describing the bulk properties of nuclear matter
as well as the properties of finite nuclei\cite{QHD}.
To incorporate the substructure of the nucleon in a relativistic nuclear 
framework, Guichon proposed a successful hybrid model 
(the quark-meson coupling (QMC) model)~\cite{Guichon88},
where nuclear matter is described in terms of non-overlapping, 
MIT bag nucleons~\cite{MIT}. 
The model was later developed to describe finite nuclei\cite{QMC96} 
as well as the properties of other hadrons in medium~\cite{QMCeta}.
By analogy with QHD,  QMC describes the bulk properties 
of nuclear systems using scalar $(\sigma)$  and vector $(\omega)$ meson 
mean fields. 
However, the nucleon bound in a nuclear medium here 
is no longer a point-like particle, 
it has  substructure -- quarks confined inside the nucleon bag.
It is the quark, rather than the nucleon itself, that is coupled to
the $\sigma$ and $\omega$ fields directly. 
As a result, the internal structure of the bound nucleon 
is modified by the surrounding medium with respect to that in free 
space~\cite{Tony94,BM96,QMChina}.

Within QMC, the small mass of the quark implies that the lower  
component of the quark wave function will be 
enhanced rapidly~\footnote{Considerably more rapidly than that of the 
nucleon in QHD and for smaller values of the mean-fields.}
by the  
change of its environment (as the $\sigma$ field strength increases),
with a consequent decrease in the scalar baryon density. 
As the scalar baryon density itself is the source of the $\sigma$ field, 
this provides a mechanism for the saturation of nuclear matter 
where the quark substructure plays a vital role.
The extra degrees of freedom, corresponding to the internal structure of 
the nucleon, lead to a reasonable value for the nuclear incompressibilty, 
once the corresponding quark and meson coupling constants, 
$g^q_\sigma$ and $g^q_\omega$, are determined to reproduce 
the empirical values for the saturation density and binding energy 
of symmetric nuclear matter. 

In our previous work, we predicted medium modifications of nucleon 
electromagnetic form factors~\cite{Lu99}. 
Such a medium effect appears to be supported by a recent experiment 
at Mainz~\cite{Dieterich01}, 
which measured the polarization transfer in the $^4He(\vec{e},e'\vec{p})$ 
reaction. The polarization transfer double ratio, 
$(P'_x/P'_z)_{\rm He} / (P'_x/P'_z)_{\rm free}$, 
tends to favor the RDWIA calculations using 
a medium modified proton form factor. 
Further study of this reaction has been carried out at 
Jefferson Lab (E93-049)~\cite{JLab01} 
and other related experiments have been proposed.
In this paper, we study the axial form factor of the bound nucleon
in symmetric nuclear matter. This is of particular interest in the light of
plans to build very high intensity neutrino beams in the near future.

%\section{Aspects of the generalized QMC model}

The Lagrangian density for the QMC model is 
\begin{eqnarray}
 \protect{\cal L}_q
  &=& \overline q (i\gamma^\mu \partial_\mu - m_q) q\theta_V - B_0\theta_V
 + g_\sigma^q \overline q q \sigma - g_\omega^q \overline q \gamma_\mu 
 q \omega^\mu - {1\over 2} m_{\sigma}^2 \sigma^2 + {1\over 2}
 m_{\omega}^2 \omega^2,
\end{eqnarray}
where $m_q$ is the current quark mass, $B_0$ is the bag constant 
in vacuum, $g_\sigma^q$ and $g_\omega^q$ 
are the corresponding quark and meson coupling constants, 
and $\theta_V$ is a step function which is one inside the bag volume
and zero outside.

In mean-field-approximation, the meson fields are treated as 
classical fields, 
and the quark field $q(x)$ inside the bag satisfies the equation of motion
\begin{equation}
[i\gamma^\mu\partial_\mu - (m_q - g_\sigma^q\overline{\sigma})-
g_\omega^q\overline{\omega} \gamma^0] q(x) = 0,
\end{equation}
where $\overline{\sigma}$ and $\overline{\omega}$ denote the constant 
mean values of the scalar and the time component of the vector field 
in symmetric nuclear matter. The normalized solution for the 
lowest state of the quark is given by\cite{Tony94,MIT} 
\begin{equation}
q(t,\vec{r}) =  {N_0\over\sqrt{4\pi}}  e^{-i\epsilon_q t/R} \left (
\begin{array}{c} g(r) \\ 
i \bfgreek {\sigma} \cdot\hat{\bf r} f(r) \end{array} \right )
\, \theta(R-r)\, \chi_q \, , \label{cavity}
\end{equation}
where 
\begin{eqnarray}
g(r) &=& j_0(x r/R), \hspace{1cm}
f(r) = \beta_q j_1(x r/R)\, , \nonumber \\
\epsilon_q &=& \Omega_q + g_\omega^q\overline{\omega} R, \hspace{1cm}
\beta_q = \sqrt{\Omega_q-m_q^*R \over \Omega_q + m_q^*R}\, , \nonumber \\
N_0^{-2} &=& 2R^3 j_0^2(x)[\Omega_q(\Omega_q-1)+ m_q^*R/2]/x^2\, , \nonumber
\end{eqnarray}
with $\Omega_q \equiv \sqrt{x^2 + (m_q^*R)^2}$, 
$m_q^* \equiv m_q - g_\sigma^q\overline{\sigma}$, 
$R$ ($R_0$) the bag radius (free space), 
and $\chi_q$ the quark Pauli spinor. The eigen-frequency, $x\, (x_0)$, 
of this lowest mode in medium (free space)
is determined by the boundary condition at the bag surface,
\begin{equation}
j_0(x) = \beta_q j_1(x).
\end{equation}
The form of the quark wave function in Eq.~(\ref{cavity}) 
is almost identical to that of the solution in free space. 
However the parameters in the expression
have to be substantially modified by the surrounding nuclear medium. 
Note that as the value of $g^q_\sigma \overline{\sigma}$  
is usually much larger than $m_q$, the quantity, $\beta_q$, 
becomes larger than unity, which means that the lower component of the
Dirac spinor is enhanced. 
In other words, the quarks in the nucleon embedded in the nuclear medium
are more relativistic than those in a  free nucleon. 

The mean values of the scalar ($\overline{\sigma}$)
and vector ($\overline{\omega}$) fields in symmetric nuclear matter 
are self-consistently determined by solving 
the following set of equations:
\begin{eqnarray}
\overline{\omega} &=& {g_\omega\rho\over m_\omega^2}, \label{omega}\\
\overline{\sigma} &=& {g_\sigma\over m_\sigma^2} 
C(\overline{\sigma}) \rho_s 
= {g_\sigma\over m_\sigma^2} 
C(\overline{\sigma}){4\over (2\pi)^3}
\int^{k_F} d^3 k {m_N^*(\overline{\sigma})\over 
\sqrt{m_N^{*2}(\overline{\sigma}) + k^2}}, \label{scc}\\
m_N^*(\overline{\sigma}) &=& {3\Omega_q (\overline{\sigma}) \over R}
- { z_0\over R} + {4\over 3}\pi R^3 B_0, \label{mass}\\
{\partial m_N^*(\overline{\sigma}) \over \partial R} &=& 0, \label{equil}
\end{eqnarray}
where $\rho$ ($\rho_s$) is the baryon (scalar) density and 
$k_F$ is the nucleon Fermi momentum, $g_\sigma= 3g_\sigma^qS(0)$,
$g_\omega= 3g_\omega^q$, and 
the quantity, $C(\overline{\sigma})$, is defined by 
\begin{equation}
C(\overline{\sigma}) \equiv S(\overline{\sigma})/S(0)=
-\left({\partial m_N^*(\overline{\sigma})
\over\partial\overline{\sigma}}\right)/g_\sigma,
\end{equation}
with $S(\overline{\sigma}) = \int_{\rm bag} d^3r\, 
\overline{q}(\vec{r})q(\vec{r})$.
Using the  quark wave function for the MIT bag, Eq.(\ref{cavity}),  
$S(\overline{\sigma})$ can be explicitly evaluated: 
\begin{equation}
S(\overline{\sigma}) = 
[\Omega_q/2+m_q^*R(\Omega_q-1)]/[\Omega_q(\Omega_q-1)+m_q^*R/2]. \label{ssigma}
\end{equation}
The second term in Eq.(\ref{mass}), ($-z_0/R$), has multiple roles and 
it parametrizes the sum of the 
zero point energy,  gluon corrections, and the part of the 
center-of-mass (c.m.) motion\cite{Guichon96}.

The process to solve this coupled system is as follows: 
we first determine $z_0$ and $B_0$ by requiring 
the free nucleon  mass to be $m_N(\rho=0) = 939 $ MeV  and by
imposing the stability condition, Eq.(\ref{equil}),
for a given bag radius, $R_0$ 
(treated as an input parameter).
After that, we solve the coupled set of equations at normal nuclear matter 
density, $\rho_0$, and determine
the coupling constants, $g_\sigma^q$ and $g_\omega^q$, required 
to reproduce nuclear saturation properties.
With these parameters, we can solve the whole problem
for each finite nuclear matter density, $\rho$, self-consistently. 
Typically the quark r.m.s. 
radius, $r^*_q$, calculated by the bag wave function is slightly increased,  
although the bag radius, $R$, decreases by a few percent 
at normal nuclear matter density. 
The properties of these self-consistent solutions as a function of
the nuclear matter density,
a typical set of parameters and the value of the nuclear 
incompressibility, $K$, as well as the possible medium dependence of 
the bag constant, can be seen in Ref.~\cite{Lu99}.

It is worthwhile to note that the self-consistency condition in QMC is 
identical to that in QHD, except that in 
QHD one has $C(\overline{\sigma})=1$ in Eq.(\ref{scc})~\cite{Tony94}, 
which corresponds to a point-like nucleon.
Within QMC all information on the internal structure of the nucleon 
is contained in $C(\overline{\sigma})$.

%\section{electromagnetic radius of the nucleon}

Once the quark wave function in the bound nucleon is determined, 
one can proceed to calculate the nucleon axial form factors,
which are defined as follows:
\begin{equation}
\bra {p' s'} A^\mu_a(0) \ket {p s} 
= \overline{u}_{s'}(p')\left[G_A(Q^2)\gamma^\mu + 
 {G_P(Q^2)\over 2m_N} (p'-p)^\mu\right]\gamma_5 {{\bf \tau}_a \over 2} u_s(p),
\end{equation}
where $Q^2 \equiv -(p'-p)^2$, $u_s(p)$ is the nucleon Dirac spinor.
Here we shall focus on the axial vector form factor, $G_A(Q^2)$, 
since the induced pseudoscalar form factor, $G_P(Q^2)$,
is dominated by the pion pole and thus can be derived using
the familiar PCAC relation~\cite{Tegen83,CBM}.
The relevant axial current operator is then simply
\begin{eqnarray}
A^\mu_a(x) &=& \sum_f \overline{q}_f(x) \gamma^\mu\gamma_5
{\tau_a\over 2} q_f(x) \theta(R-r), 
\label{current}
\end{eqnarray}
where $q_f(x)$ is the quark field operator for flavor $f$.

In order to remove the spurious c.m. motion, we construct 
the momentum eigenstate of a baryon  
via the Peierls-Thouless (PT) projection method\cite{Lu97,PT62},
\begin{equation}
\Psi_{\rm{PT}}(\vec{x}_1, \vec{x}_2, \vec{x}_3; \vec{p}) = 
N_{\rm{PT}} e^{i\vec{p} \cdot \vec{x}_{\rm{c.m.}} }
q(\vec{x}_1 - \vec{x}_{\rm{c.m.}}) q(\vec{x}_2 - \vec{x}_{\rm{c.m.}})
q(\vec{x}_3 - \vec{x}_{\rm{c.m.}}), \label{PTWF}
\end{equation}
where $N_{\rm{PT}}$ is a normalization  constant, 
$\vec{p}$  the total momentum of the baryon, and 
$\vec{x}_{\rm{c.m.}} = (\vec{x}_1 + \vec{x}_2 + \vec{x}_3)/3$ 
is the center of mass of the baryon (we assume equal mass quarks here).
It can be shown that the PT wave function satisfies the condition of
translational invariance.
Using Eqs.~(\ref{current}) and (\ref{PTWF}), the nucleon axial
form factors can be expressed as
\begin{eqnarray}
G_A(Q^2) &=& {5\over 3}\int\! d^3r \left\{\left[g^2(r)-f^2(r)\right]j_0(Qr) 
  +2f^2(r){j_1(Qr)\over Qr}\right\} K(r)/D_{\rm PT}, \label{PTM}\\
D_{\rm PT} &=& \int\! d^3r \rho_q(r) K(r),
\end{eqnarray}
where 
$D_{\rm PT}$ is the normalization factor,
$\rho_q(r) \equiv g^2(r) + f^2(r)$, and 
$K(r) = \int\! d^3z \, \rho_q(\vec{z}) \rho_q(-\vec{z} - \vec{r})$
is the recoil function which accounts for the correlation of the 
two spectator quarks.

At this stage, there is no satisfactory covariant treatment for the MIT bag 
model. On the other hand, relativistic effects are important for
most dynamic variables, 
especially for form factors at large momentum transfer.
They lead to a sizable correction for the r.m.s. radius of the nucleon.
In this paper,  we use a semi-phenomenological method to account for the 
relativistic corrections which is consistent with a mean field treatment of 
the QMC model.
Since a static MIT bag is an extended spherical object, it would be deformed 
if it were viewed in a moving frame of reference.
It is crucial to include
this Lorentz contraction of the bag for calculating form factors 
at moderate momentum transfer\cite{Lu97,LP70}.
In the prefered Breit frame, the resulting form
factors can be expressed through a simple rescaling, i.e.,
\begin{eqnarray}
G_A(Q^2) &=& \left({m_N^*\over E^*}\right)^2 
        G^{\rm sph}_A(Q^2 {m_N^*}^2/{E^*}^2),
\end{eqnarray}
where $E^*=\sqrt{{m_N^*}^2 + Q^2/4}$ and 
$G_{A}^{\rm sph}(Q^2)$ are the form factors calculated 
with the  spherical bag wave function.
The scaling factor in the argument arises from the coordinate transformation
of the struck quark whereas
the prefactor, $(m_N^*/E^*)^2$,  comes from  the reduction 
of the integral measure of two spectator quarks in the Breit frame\cite{LP70}.
The axial radius squared is given by,
\begin{equation}
r^2_{A}  = -{6\over g_A} \left. 
{dG_A(Q^2)\over d  Q^2}
\right|_{Q^2\rightarrow 0}, 
\end{equation}
which results in $12/m_A^2$ for a dipole form: 
$G_A(Q^2)=1/(1+Q^2/m_A^2)^2$.

%\section{pion clouds and meson exchange currents}

Note that the pion cloud of the nucleon only plays an indirect 
role in calculating the axial vector form factor, in contrast to the 
electromagnetic properties of the nucleon. 
In fact, the matrix element 
of the pionic axial current, $f_\pi\partial^\mu \pi$, vanishes in 
any chiral bag model 
if the pion cloud exists in all space~\cite{Tegen83,Lu01todo}.  

%\section{Discussion of the Results}

\begin{figure}
\vspace{2.5cm}
\centering{\
\epsfig{file=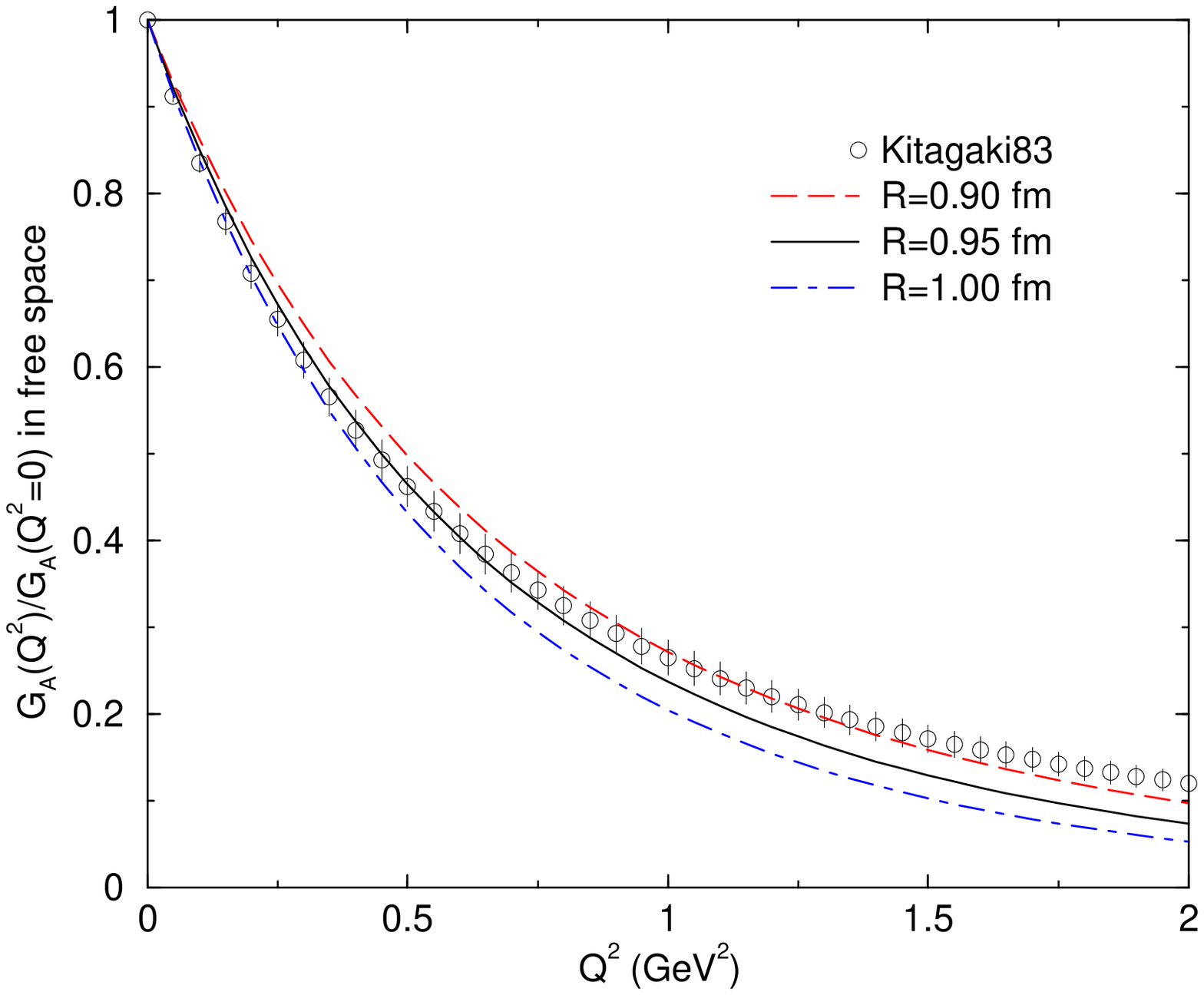,width=12cm}
%\vspace*{1cm}
%\epsfbox{fig1.ps}
\caption{The nucleon axial form factor in free space for three different 
bag radii. Experimental data are summarised by a dipole form: 
$G_A(Q^2)=g_A/(1+Q^2/m_A^2)^2$, with $m_A=(1.03\pm 0.04)$ GeV. 
The value of $g_A$ in our calculation 
is 1.14, compared with the experimental value of 1.26. }
\label{gAfig1.eps}}
\end{figure}

The nucleon axial form factor in free space is illustrated in 
Fig.~\ref{gAfig1.eps}.
The experimental data from neutrino scattering is not shown directly, rather
we show a dipole form with the mass parameter in the range found by 
Kitagachi et al.~\cite{Kitagaki83}.
(We note that the pion electroproduction data~\cite{Liesenfeld99} 
is consistent with this after the small correction from 
chiral perturbation theory is applied~\cite{Meissner01}.)
The corrections for the center-of-mass motion and Lorentz contraction
lead to a significant improvement over static bag model calculations, 
in particular, at moderate momentum transfers~\cite{Lu01todo}. 
The results are not sensitive to the current quark mass, so 
we use $m_q=0$ in this paper.
The axial radii are 0.587, 0.614 and 0.640 fm, 
corresponding to the bag radii of 0.90, 0.95 and 1.00 fm, respectively.
These r.m.s. radii are in good agreement with the experimental 
value $(0.635\pm 0.023)$ fm~\cite{Thomas_book}. 
The axial coupling constant, $g_A \equiv G_A(Q^2=0)$, is about 1.14 
after the c.m. correction, 
which is about 5\% increase from the static MIT bag value, 1.09.
Finite quark mass and pion renormalization may be expected to lead to further
corrections at the level of 10\%.

\begin{figure}
\vspace{2.5cm}
\centering{\
\epsfig{file=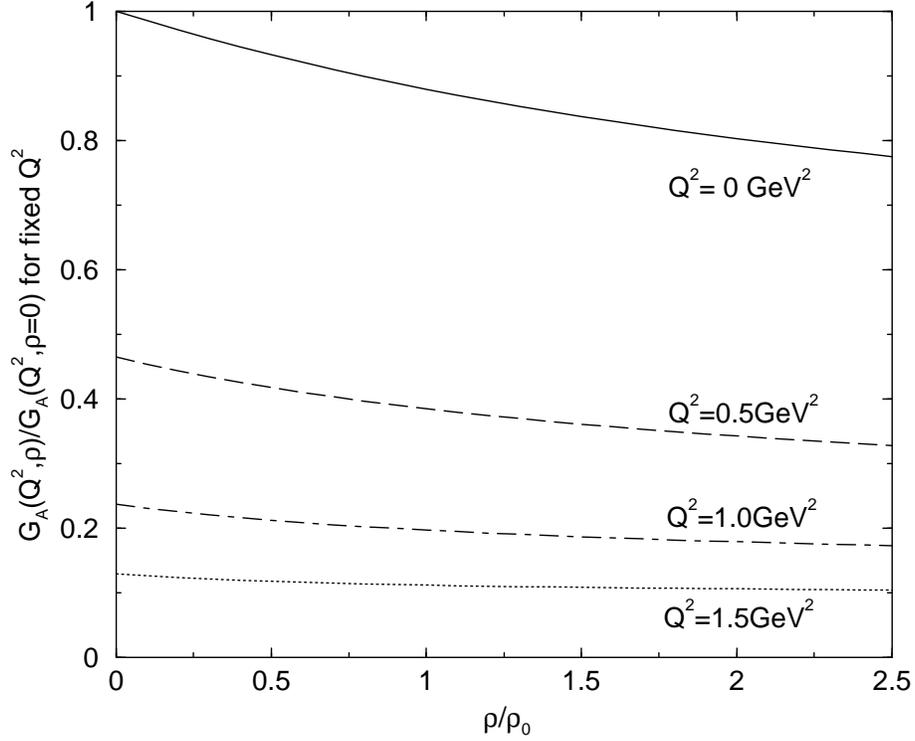,width=12cm}
%\vspace*{1cm}
%\epsfbox{fig1.ps}
\caption{The density dependence of the nucleon axial form factor
with $R_0 = 0.95$ fm. The value of $g_A$ is quenched in nuclear matter, 
resulting in a reduction of roughly 12\% and 9\% at $\rho_0$ and 0.7$\rho_0$, 
respectively.}
\label{gAfig2.eps}}
\end{figure}

Fig.~\ref{gAfig2.eps} shows the medium dependence of the axial form factors
at several momentum transfers (with R$_0$ = 0.95 fm).
For $Q^2 < 2$ GeV$^2$, the axial form factor decreases as 
the density $\rho$ increases.
The axial vector coupling constant (the solid line in this figure), $g_A$, 
would be reduced by 12\% at normal nuclear matter density, $\rho_0$, 
and 9\% at the average density of finite nuclei, 0.7$\rho_0$, 
where the latter is    
to be compared with the experimental 20\% reduction seen 
in Gamow-Teller transitions~\cite{gA}.  

\begin{figure}
\vspace{2.5cm}
\centering{\
\epsfig{file=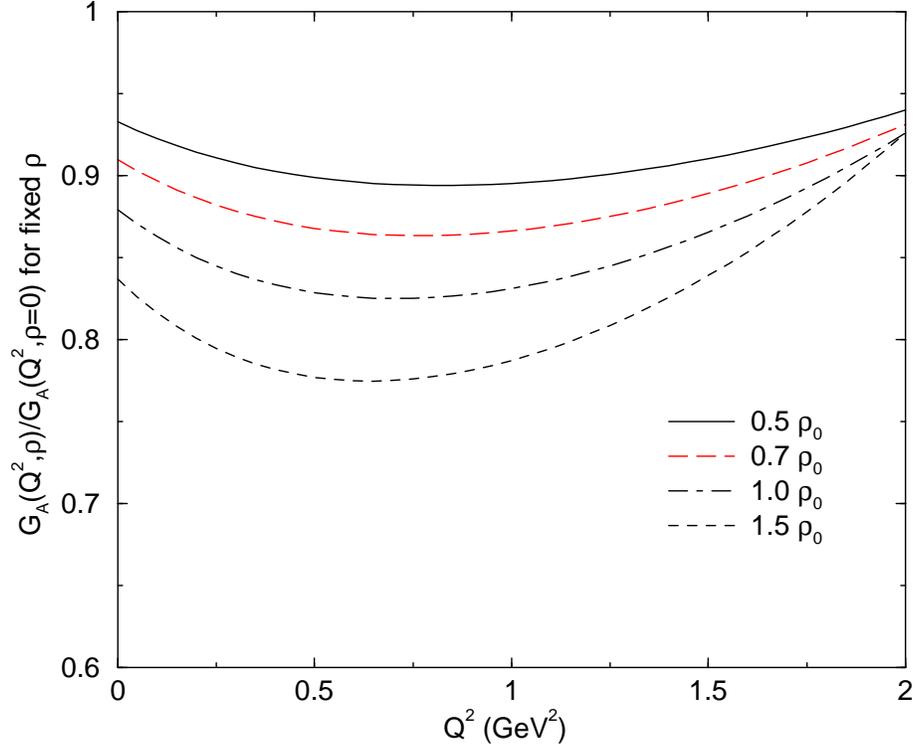,width=12cm}
%\vspace*{1cm}
%\epsfbox{fig1.ps}
\caption{The momentum dependence of the nucleon axial form factor at 
different densities with $R_0=0.95$ fm.
The effect of medium modification, which is more important in small momentum
transfer region, glows as the baryon density increases.}
\label{gAfig3.eps}}
\end{figure}

The momentum dependence of the axial form factors (R$_0$ = 0.95 fm again) 
is shown in 
Fig.~\ref{gAfig3.eps}. The nuclear medium does indeed modify 
the shape of the momentum dependence.
The medium effect increases as the density increases and
tends to be larger for smaller momentum transfers. 
However the overall effect is less than roughly 8\%, depending
on the bag radius.  
 
In conventional nuclear physics, the nucleon is immutable and 
the electroweak properties of a nucleus are often described by 
a combination of individual nucleon contributions and various  
meson exchange current corrections. 
This is particularly important for the axial charge of 
finite nuclei~\cite{Kubodera78}. 
For the quenching of $g_A$, conventional mechanisms 
(medium polarization) also exist,
however it is fundamentally interesting to explore this new mechanism 
arising from possible changes of the internal structure of the nucleon. 
Clearly it would be very valuable to study the effect of the medium modified 
axial form factor in the context of 
neutrino-nucleus scattering\cite{Kubodera94} and 
the solar neutrino problem~\cite{Haxton95}.

%\section{Conclusion}

In summary, the nucleon axial form factor is significantly 
modified in nuclear matter through a new mechanism, namely the change 
of the internal substructure of the 
bound nucleon. 
We have calculated the density dependence of the 
axial form factor of the bound nucleon in the QMC model and 
found that it is 
quenched considerably compared with that of the free nucleon.
The axial vector coupling constant, $g_A$, is reduced by roughly 10\% 
at $\rho_0$ and the axial form factor would vary within 8\%  
for moderate momentum transfers.

%\section{Acknowledgements}

D.H.Lu is grateful to the Y.C.Tang Disciplinary Development Fund in 
Zhejiang University and would like to acknowledge the warm hospitality 
of the CSSM in Adelaide University. 
This work was supported in part by the National Natural Science Fund of China 
and by the Australian Research Council.

\end{document}